\documentclass[fleqn,10pt]{wlscirep}
\title{Topological spin and valley pumping in silicene}
\author[1]{Wei Luo}
\author[1,2,*]{L. Sheng}
\author[1,2]{B. G. Wang}
\author[1,2,$\dagger$]{D. Y. Xing}
\affil[1]{National Laboratory of Solid State Microstructures and Department of Physics, Nanjing University, Nanjing 210093, China}
\affil[2]{Collaborative Innovation Center of Advanced Microstructures,
Nanjing University, Nanjing 210093, China}
\affil[*]{shengli@nju.edu.cn}
\affil[$\dagger$]{dyxing@nju.edu.cn}

\begin{abstract}
We propose to realize adiabatic topological spin and
valley pumping by using silicene, subject to the modulation  
of an in-plane $ac$ electric field
with amplitude $E_{y}$ and a vertical electric field consisting of an
electrostatic component and an $ac$ component with amplitudes $E_{z}^{0}$
and $E_{z}^{1}$. By tuning $E_{z}^{0}$ and $E_{z}^{1}$, topological valley
pumping or spin-valley pumping can be achieved. The low-noise valley and
spin currents generated can be useful in valleytronic and spintronic
applications. Our work also demonstrates that bulk topological spin or valley
pumping is a general characteristic effect of two-dimensional topological
insulators, irrelevant to the edge state physics.
\end{abstract}
\begin{document}

\maketitle

\section*{Introduction}

Topological transport phenomena are generally protected by certain topological
invariants, and exhibit universal properties that are immune to impurity
scattering and insensitive to material details. Since the discovery of the
integer quantum Hall (IQH) effect in two-dimensional (2D) electron
systems~\cite{kli} in 1980, the first example of the topological transport
phenomena, the fascinating characteristics of topological transport continue
to be the primary focus of more and more research activities. Laughlin
interpreted the precise integer quantization of the Hall conductivity in units
of $e^{2}/h$ in the IQH effect in terms of an adiabatic quantum charge
pump~\cite{laugh}. Thouless, Kohmoto, Nightingale, and Nijs established a
relation between the quantized Hall conductivity and a topological
invariant~\cite{thou}, namely, the TKNN number or the Chern number. Thouless
and Niu further related the amount of charge pumped in a charge pump to the
Chern number~\cite{niu}.

In recent years, the quantum spin Hall (QSH) effect, a spin analogue of the
IQH effect, was proposed theoretically~\cite{kane,bern}, and realized
experimentally in $\mathrm{HgTe}$ quantum wells~\cite{konig} and
$\mathrm{InAs/GaSb}$ bilayers~\cite{knez}. A QSH system, also called a 2D
topological insulator (TI)~\cite{has,qi}, is insulating in the bulk with a
pair of gapless helical edge states at the sample boundary. In the ideal case,
where the electron spin is conserved, a QSH system can be viewed as two
independent IQH systems without Landau levels~\cite{hald}, so that the
topological properties of the system can be described by the opposite Chern
numbers of the two spin species. In general, when the electron spin is not
conserved, unconventional topological invariants, either the $Z_{2}$
index~\cite{kane1} or spin Chern numbers~\cite{sheng,pro,yang}, are needed to
describe the QSH systems. The time-reversal symmetry is considered to be a
prerequisite for the QSH effect, which protects both the $Z_{2}$ index and
gapless nature of the edge states. However, based upon the spin Chern
numbers, it was shown that the bulk topological properties remain intact
even when the time-reversal symmetry is broken. This finding evokes the
interest to pursue direct investigation and utilization of the robust
topological properties of the TIs, besides using their symmetry-protected
gapless edge states, which are more fragile in realistic environments.

Recently, Chen $et$ $al.$ proposed that a spin Chern pumping effect from the
bulk of the 2D TI, a $\mathrm{HgTe}$ quantum well, can be realized by using
time-dependent dual gate voltages and an in-plane $ac$ electric field~\cite{chen},
which paves a way for direct investigation and utilization of the bulk
topological properties of the TIs. The work of Chen $et$ $al.$ is a
generalization of the earlier proposals of topological spin pumps~\cite
{Sharma,Shindou,Z2pump,TopoClassification,CQZhou}, based upon 1D abstract
models, to a realistic 2D TI material. The spin Chern pump is a full
spin analogue to the Thouless charge pump, in the sense that it is 
driven by topological invariants alone, without relying on any symmetries.
For example, it has been shown that magnetic impurities breaking both
spin conservation and time-reversal symmetry only modify the amount 
of spin pumped per cycle in a perturbative manner~\cite{chen,CQZhou}, being essentially
distinct from the QSH effect. Wan and Fischer suggested to realize
a topological valley resonance effect in graphene by using the
time-dependent lattice vibration of optical phonon modes, which can pump out
a noiseless and quantized valley current flowing into graphene leads~\cite%
{JunWan}. This topological valley resonance effect is intimately related to
the spin or valley Chern pumping, as it is solely attributable to the valley
Chern numbers, independent of the time-reversal symmetry~\cite{JunWan}.

Silicene, the cousin of graphene, is a monolayer of silicon atoms instead of
carbon atoms on a 2D honeycomb lattice. Recently, this material has been
experimentally synthesized~\cite{lalmi,vogt,lin} and theoretically explored~%
\cite{cah,liu,ezawa,pan}. Similar to graphene, the energy 
spectrum of silicene has two Dirac valleys, around the $K
$ and $K^{\prime }$ points sited at opposite corners 
of the hexagonal Brillouin zone. 
Silicene has a much larger spin-orbit gap than graphene, favoring the QSH
effect. As another prominent property distinguishing it from graphene,
silicene has a buckled lattice structure, which allows us to 
control the Dirac masses  at $K
$ and $K^{\prime }$ points independently, by applying an external 
vertical electric field~\cite{ezawa,missa}. This property
also makes silicene be a natural candidate for valleytronics~\cite
{rycerz,xiao,akhm}.

In this paper, we propose an experimental scheme
to achieve topological spin and valley pumping by
applying in silicene an in-plane $ac$ electric field with amplitude $E_{y}$
and a vertical electric field comprising an electrostatic component and an $
ac$ component  with amplitudes $E_{z}^{0}$ and $E_{z}^{1}$. The present
proposal is more  practicable experimentally 
than the previous one~\cite{chen}, because
applying a vertical electric field in silicene has been much better
understood~\cite{ezawa,missa} and is more practical than applying dual gate
voltages in HgTe quantum wells. By using the spin-valley Chern numbers, it
is shown that the system can be in the pure valley pumping regime, mixed spin
and valley pumping regime, or trivial pumping regime, depending on the strengths
$E_{z}^{0}$ and $E_{z}^{1}$ of the perpendicular electric field. The total
amount of valley or spin quanta pumped per cycle, calculated from
the scattering matrix formula, is fully consistent with the spin-valley
Chern number description. It is proportional to the
cross-section of the sample, and insensitive to the material parameters, a
clear evidence that the pumping is a bulk topological effect, irrelevant to
the edge states.


\section*{Results}

\begin{figure}[tpb]
\begin{center}
\includegraphics[width=5.0in,keepaspectratio]{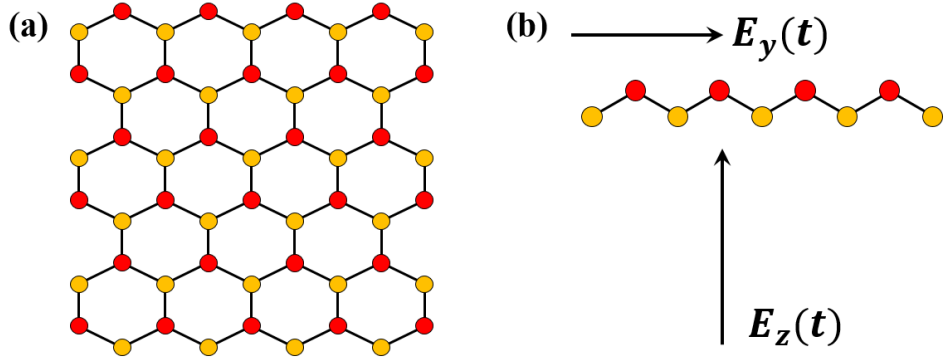}
\caption{(a) The honeycomb lattice and (b) buckled structure of silicene.
$E_{y}(t)$ and $E_{z}(t)$ are the time-dependent electric fields along the $y$ and $z$
directions, respectively.}\label{Fig.3}
\end{center}
\end{figure}

{\bf Model Hamiltoinan.} Silicene consists of a honeycomb lattice of silicon
atoms with two sublattices of $A$ and $B$ sites, as shown in Fig. 1. We consider
a silicene sheet in parallel to the $xy$ plane. Different from graphene, silicene
has a buckled structure, i.e., the two sublattice planes are separated by a small
distance $l\simeq 0.44{\mathrm{\mathring{A}}}$ along the $z$ direction~\cite{cah}.
Silicene can be described by the tight-binding model
\begin{equation}
H_{0}=-t\sum_{\langle i,j\rangle\sigma}c_{i\sigma}^{\dagger}c_{j\sigma} +
i\frac{\lambda_{\mathrm{SO}}}{3\sqrt{3}}\sum_{\langle\langle
i,j\rangle\rangle\sigma\sigma^{\prime }}
v_{ij}c_{i\sigma}^{\dagger}\sigma_{\sigma\sigma^{\prime
}}^{z}c_{j\sigma^{\prime }}\ ,  \label{H_TI}
\end{equation}
where $c_{i\sigma}^{\dagger}$ creates an electron with spin polarization
$\sigma=\uparrow$ or $\downarrow$ at site $i$, and $\langle i,j\rangle$ and
$\langle\langle i,j\rangle\rangle$ run over all the nearest-neighbor and
next-nearest-neighbor sites. The first term describes the nearest-neighbor
hopping of the electrons with $t=1.6 \mathrm{eV}$. The second term represents
the intrinsic spin-orbit coupling with $\lambda_{\mathrm{SO}}=3.9 \mathrm{meV}$,
where $v_{ij}=1$ if the next-nearest-neighbor hopping is counterclockwise
around a hexagon with respect to the positive $z$ axis, and $v_{ij}=-1$ if
the hopping is clockwise.

For the following calculations, it is sufficient to use the low-energy
continuum Hamiltonian, which can be obtained by expanding Hamiltonian (1)
around the Dirac points $K$ and $K^{\prime }$ to the linear order in the
relative momentum
\begin{equation}
H_{0}=v_{\mbox{\tiny F}}(\eta k_{x}\hat{\tau}_{x}+k_{y}\hat{\tau}_{y})
+\lambda_{\mathrm{SO}}\eta\hat{\tau}_{z}\hat{\sigma}_{z}\ ,  \label{H_TI}
\end{equation}
where $\mathbf{k}=(k_x,k_y)$ is the relative momentum, $\eta=\pm$ correspond
to the $K$ and $K^{\prime }$ valleys, and $v_{\mbox{\tiny F}}=\frac{\sqrt{3}}{2}at$
is the Fermi velocity with the lattice constant $a=3.86\mathrm{{\mathring{A}}}$.
To drive the quantum pumping, two time-dependent electric fields are applied to
the system. One is along the $z$ direction of the form
$E_{z}(t)=E_{z}^{0}+E_{z}^{1}\cos{(\omega t)}$ with $E_{z}^{0}$ and $E_{z}^{1}$ as
the amplitudes of the electrostatic component and $ac$
component, respectively. The other is an $ac$ electric field along the
negative $y$ direction, $E_{y}(t)=-E^{0}_{y}\cos{(\omega t)}$. By taking the
two electric fields into account, the Hamiltonian is rewritten as
\begin{equation}
H_{\mathrm{P}}=v_{\mbox {\tiny F}}[\eta k_{x}\hat{\tau}_{x}+(k_{y}-eA(t))\hat{\tau}_{y}] +\lambda_{\mathrm{SO}}\eta\hat{\tau}_{z}\hat{\sigma}_{z}
-l(E_{z}^{0}+E_{z}^{1}\cos{(\omega t)})\hat{\tau}_{z}\ .  \label{H_TI}
\end{equation}
Here $-e$ is the electron charge, and $A(t)=A_{y}\sin{(\omega t)}$ is the
vector potential of the $ac$ electric field along the negative $y$ direction
with $A_{y}=E_{y}/\omega$ and $\omega>0$ being assumed.

\begin{figure}[tpb]
\begin{center}
\includegraphics[width=5.0in,keepaspectratio]{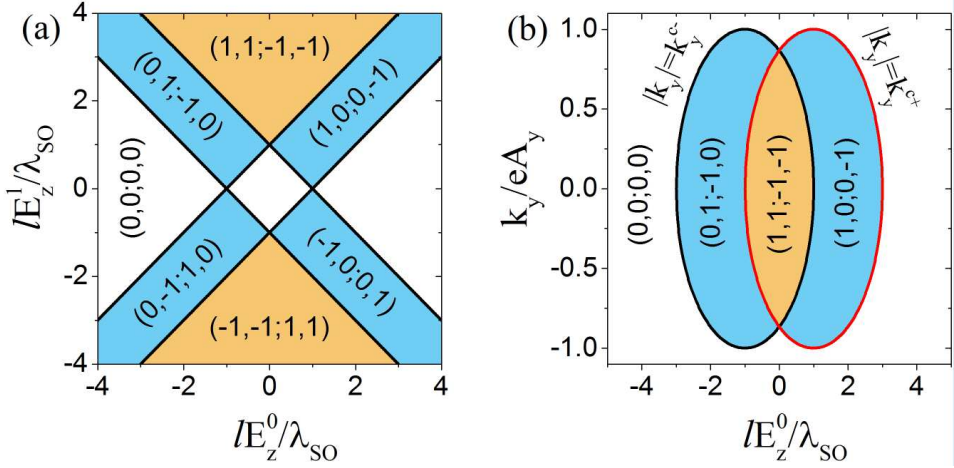}
\end{center}
\caption{(a) The phase diagram of the spin-valley Chern numbers on the
normalized $E_{z}^{1}$ vs normalized $E_{z}^{0}$ plane for $k_{y}=0$, and
(b) the phase diagram on the $k_{y}/eA_{y}$ vs normalized $E_{z}^{0}$ plane
for $lE_{z}^{1}=2\protect\lambda_{\mathrm{SO}}$. The numbers in the brackets
are spin-valley Chern numbers, i.e., $(C_{+}^{\uparrow},C_{+}^{%
\downarrow};C_{-}^{\uparrow},C_{-}^{\downarrow})$. $E_{y}$ is taken to be
positive, and for negative $E_{y}$, all the spin-valley Chern numbers in the
phase diagrams will flip signs. }
\label{Fig.2}
\end{figure}

{\bf Spin-valley Chern numbers.} Within the adiabatic approximation, for a bulk sample,
one can obtain for the eigenenergies of Eq.\ (\ref{H_TI}) at any given time $t$
\begin{eqnarray}
E(\mathbf{k})=\Bigl\{v^2_{\mbox{\tiny F}} k_{x}^2+v^2_{\mbox{\tiny F}}
\bigl(k_{y}-eA(t)\bigr)^2
+\bigl[\lambda_{\mathrm{SO}}\eta\xi_{\sigma}- l(E_{z}^{0}+E_{z}^{1}
\cos{(\omega t)})\bigr]^2\Bigr\}^{1/2}\ ,
\end{eqnarray}
where $\xi_{\uparrow}=-\xi_{\downarrow}=1$. We note that $E(\mathbf{k})$ depends on
valley $\eta$ and spin $\sigma$ only through the product $\eta\xi_{\sigma}$, which
has two possible values, $\eta\xi_{\sigma}=\pm 1$. It is convenient to consider the
whole system as consisting of two subsystems, one with $\eta\xi_{\sigma}=1$ and the
other with $\eta\xi_{\sigma}=-1$. For the $\eta\xi_{\sigma}=1$ subsystem (i.e.,
$\eta = +$ and $\sigma=\uparrow$, or $\eta=-$ and $\sigma=\downarrow$), if
$|lE_{z}^{1}|<|lE_{z}^{0}-\lambda_{\mathrm{SO}}|$, there always exists a
finite energy gap between the conduction and valence bands. If
$|lE_{z}^{1}|\ge|lE_{z}^{0}-\lambda_{\mathrm{SO}}|$, at
$\cos{(\omega t)}=\frac{\lambda_{\mathrm{SO}}-lE_{z}^{0}}{lE_{z}^{1}}$, the conduction
and valence bands of the subsystem touch at $k_{x}=0$ and $k_{y}=k_{y}^{c+}$ or $-k_{y}^{c+}$. Similarly, for the $\eta\xi_{\sigma}=-1$ subsystem ($\eta = +$ and $\sigma=\downarrow$, or
$\eta=-$ and $\sigma=\uparrow$), if $|lE_{z}^{1}|<|lE_{z}^{0}+\lambda_{\mathrm{SO}}|$,
there always exists a finite energy gap between the conduction and valence bands. If
$|lE_{z}^{1}|\ge|lE_{z}^{0}+\lambda_{\mathrm{SO}}|$, at
$\cos{(\omega t)}=\frac{-\lambda_{\mathrm{SO}}-lE_{z}^{0}}{lE_{z}^{1}}$, the conduction
and valence bands touch at $k_{x}=0$ and $k_{y}=k_{y}^{c-}$ or $-k_{y}^{c-}$.
Here,
\begin{equation}
k_{y}^{c\pm}=\left|eA_{y}\frac{\sqrt{(lE_{z}^{1})^{2}-
(lE_{z}^{0}\mp\lambda_{\mathrm{SO}})^{2}}}{lE_{z}^{1}}\right|\ .
\end{equation}

It has been established that the nontrivial topological properties of the system
accounting for the spin or valley pumping can be well described by the spin-valley
Chern numbers~\cite{chen}. The topological pumping can be visualized as the quantized
spectral flow of the spin-polarized Wannier functions, which originates from the
nonzero spin-valley Chern numbers~\cite{chen}. Following Chen $et$ $al.$~\cite{chen},
we consider $k_{y}$ as a parameter, and calculate the spin-valley Chern numbers
$C_{\eta}^{\sigma}(k_y)$ in the standard way~\cite{pro,yang}, on the torus of the two
variables $k_{x}\in(-\infty,\infty)$ and $t\in[0,T)$ with $T=2\pi/\omega$ as
the period. The spin-valley Chern numbers are obtained as
\begin{eqnarray}
C_{\eta}^{\sigma}(k_y)=\eta\theta(|lE_{z}^{1}|-|lE_{z}^{0}-\lambda_{\mathrm{SO}}|)
\theta(k_{y}^{c+}-|k_{y}|)\mathrm{sgn}(E_{y}E_{z}^{1})\ ,
\label{spin-valley1}
\end{eqnarray}
for $\eta\xi_{\sigma}=1$, and
\begin{eqnarray}
C_{\eta}^{\sigma}(k_y)=\eta\theta(|lE_{z}^{1}|-|lE_{z}^{0}+\lambda_{\mathrm{SO}}|)
\theta(k_{y}^{c-}-|k_{y}|)\mathrm{sgn}(E_{y}E_{z}^{1})\ ,
\label{spin-valley2}
\end{eqnarray}
for $\eta\xi_{\sigma}=-1$, where $\theta(x)$ is the unit step function.

The phase diagram of the spin-valley Chern numbers for $k_{y}=0$ on the $E_{z}^{1}$
versus $E_{z}^{0}$ plane is plotted in Fig.\ 2(a). This phase diagram is mainly
determined by the first $\theta$-function in Eq.\ (\ref{spin-valley1}) or
(\ref{spin-valley2}), which sets four straight lines as the phase boundaries, and the
second $\theta$-function can be considered to be always equal to unity for $k_y=0$.
The yellow region can be described by the inequations
$|lE_{z}^{1}|>|lE_{z}^{0}-\lambda_{\mathrm{SO}}|$ and
$|lE_{z}^{1}|>|lE_{z}^{0}+\lambda_{\mathrm{SO}}|$. The blue region is given by
$|lE_{z}^{1}|>|lE_{z}^{0}-\lambda_{\mathrm{SO}}|$ and
$|lE_{z}^{1}|<|lE_{z}^{0}+\lambda_{\mathrm{SO}}|$, or
$|lE_{z}^{1}|<|lE_{z}^{0}-\lambda_{\mathrm{SO}}|$ and
$|lE_{z}^{1}|>|lE_{z}^{0}+\lambda_{\mathrm{SO}}|$. The white region corresponds to
$|lE_{z}^{1}|<|lE_{z}^{0}-\lambda_{\mathrm{SO}}|$ and
$|lE_{z}^{1}|<|lE_{z}^{0}+\lambda_{\mathrm{SO}}|$. One may notice that
on any of the phase boundaries, the band gap always closes at certain time.

A typical phase diagram on the $k_{y}$ versus $E_{z}^{0}$ plane for
$lE_{z}^{1}=2\lambda_{\mathrm{SO}}$ is plotted in Fig.\ 2(b), where $E_{y}$
is taken to be positive. The phase diagram can be understood as the
superposition of those of the two subsystems of $\eta\xi_{\sigma}=1$ and $-1$,
as indicated by Eqs.\ (\ref{spin-valley1}) and (\ref{spin-valley2}). The
phase diagram of the $\eta\xi_{\sigma}=1$ subsystem is determined by the
boundary $k_y=k_{y}^{c+}$, which can be rewritten into the standard form of
an ellipse equation
$k_y^{2}/(eA_{y})^2+(lE_{z}^{0}-\lambda_{\mathrm{SO}})^2/(lE_{z}^{1})^{2}=1$,
centered at $lE_{z}^{0}/\lambda_{\mathrm{SO}}=1$ and $k_{y}=0$. The spin-valley
Chern numbers of the subsystem take values
$C_{+}^{\uparrow}(k_y)= -C_{-}^{\downarrow}(k_{y})=\mathrm{sgn}(E_{y}E_{z}^{1})$
inside the ellipse, and vanish outside the ellipse. Similarly, the phase diagram
of the $\eta\xi_{\sigma}=-1$ subsystem is determined by the boundary
$k_y=k_{y}^{c-}$, which can be rewritten into the standard form of an ellipse
equation $k_y^{2}/(eA_{y})^2+(lE_{z}^{0}+\lambda_{\mathrm{SO}})^2/(lE_{z}^{1})^{2}=1$,
centered at $lE_{z}^{0}/\lambda_{\mathrm{SO}}=-1$ and $k_{y}=0$. The spin-valley
Chern numbers of the subsystem take values
$C_{+}^{\downarrow}(k_y)=-C_{-}^{\uparrow}(k_{y})=\mathrm{sgn}(E_{y}E_{z}^{1})$
inside the ellipse, and vanish outside the ellipse.

For the convenience to relate the above phase diagram to the spin and valley
pumping, we introduce the total valley Chern number
$C_{\mathrm{valley}}(k_{y})=\sum_{\eta\sigma}\eta C_{\eta}^{\sigma}(k_{y})$ and
total spin Chern number
$C_{\mathrm{spin}}(k_{y})=\sum_{\eta\sigma}\xi_{\sigma}C_{\eta}^{\sigma}(k_{y})$.
The total charge Chern number $\sum_{\eta\sigma}C_{\eta}^{\sigma}(k_{y})$ always
vanishes and will not be considered. For definiteness, we focus on the case where
$E_{z}^{1}>0$, corresponding to the upper half of the phase diagram Fig.\ 2(a). The
opposite case where $E_{z}^{1}<0$, corresponding to the lower half phase
diagram, can be understood similarly. When the system is in the yellow region of
Fig.\ 2(a), if $E_{z}^{0}=0$, we have $k_{y}^{c+}=k_{y}^{c-}$, and the spin-valley
Chern numbers
$(C_{+}^{\uparrow},C_{+}^{\downarrow};C_{-}^{\uparrow},C_{-}^{\downarrow})=(1,1;-1,-1)$
for $\vert k_y\vert < k_{y}^{c+}$ and $(0,0;0,0)$ for $\vert k_y\vert > k_{y}^{c+}$,
as can be seen from Fig.\ 2(b). The total valley Chern number $C_{\mathrm{valley}}(k_{y})=4$
for $\vert k_y\vert < k_{y}^{c+}$, and $0$ for $\vert k_y\vert > k_{y}^{c+}$. The
total spin Chern number $C_{\mathrm{spin}}(k_{y})=0$ for any $k_y$. The system is in
the pure valley pumping regime, without pumping spin. If $E_{z}^{0}>0$, as can be
seen from Fig.\ 2(b), the spin-valley Chern numbers take values $(1,1;-1,-1)$
for $\vert k_y\vert < k_{y}^{c-}$, $(1,0;0,-1)$ for $k_{y}^{c-}<\vert
k_y\vert < k_{y}^{c+}$, and $(0,0,0,0)$ for $\vert k_y\vert > k_{y}^{c+}$.
As a result, the electron states with $\vert k_y\vert < k_{y}^{c-}$ have
$C_{\mathrm{valley}}(k_{y})=4$ and $C_{\mathrm{spin}}(k_{y})=0$, and contribute to pure
valley pumping, similar to the case for $E_{z}^{0}=0$. The states with
$k_{y}^{c-}<\vert k_y\vert < k_{y}^{c+}$ contribute to both valley and spin pumping, and
pump an equal amount of valley and spin quanta per cycle, because
$C_{\mathrm{valley}}(k_{y})=C_{\mathrm{spin}}(k_{y})=2$ in this region. The other states
with $\vert k_y\vert > k_{y}^{c+}$ do not contribute to the pumping. Therefore, the system
as a whole is in a regime of mixed spin and valley pumping. Each cycle, the system pumps
more valley quanta than spin quanta. The case for $E_{z}^{0}<0$ can be analyzed similarly.
When the system is in the blue region of Fig.\ 2(a), by assuming $E_{z}^{0}>0$ for definity,
the spin-valley Chern numbers equal to $(1,0;0-1)$ for $\vert k_{y}\vert < k_{y}^{c+}$, and
$(0,0;0,0)$ for $\vert k_{y}\vert >k_{y}^{c+}$. The corresponding total valley Chern number
and spin Chern number are $C_{\mathrm{valley}}(k_{y})=C_{\mathrm{spin}}(k_{y})=2$ for
$\vert k_{y}\vert < k_{y}^{c+}$, and vanish for $\vert k_{y}\vert > k_{y}^{c+}$. The system
is in the spin-valley pumping regime. Different from the spin-valley pumping in the yellow
region of Fig.\ 2(a), each cycle, the system pumps an equal amount of valley and spin quanta.
When the system is in the white region of Fig.\ 2(a), the spin-valley Chern numbers all
vanish for any $k_y$, and the system is a trivial insulator.

{\bf Spin Pumping from The Scattering Matrix Formula.} The amount of spin and valley
quanta pumped per cycle can be conveniently calculated by using the scattering matrix
formula~\cite{butti,brou}. In the following, we show that the calculated result from
the scattering matrix formula is consistent with the above topological description.
The spin pumping is more interesting than valley pumping regarding practical
applications, and we will focus on the amount of spin pumped per cycle. 
The valley pumping can be studied similarly by considering an electrode with 
natural valley degrees of freedom. We consider
the pump is attached to a normal electrode, with a potential barrier in between. The
total Hamiltonian of the system is taken to be
\begin{equation}
H=\left\{
\begin{array}{ll}
H_{\mathrm{P}} & {(x<0)} \\
H_{\mathrm{E}}+V_{\mathrm{B}} & {(0<x<d)}\ . \\
H_{\mathrm{E}} & {(x>d)}
\end{array}
\right.\
\end{equation}
The Hamiltonian $H_{\mathrm{P}}$ at $x<0$ for the pump body is given by Eq.\ (\ref{H_TI}),
and the electrode is taken to be a normal metal with a 2D parabolic Hamiltonian
\begin{equation}
H_{\mathrm{E}}=-E_{0}+\frac{p^2}{2m}\ ,  \label{H_e}
\end{equation}
where $\mathbf{p}=(p_x,p_y)$ is the 2D momentum, and $E_{0}$ and $m$ are constant model
parameters. In the barrier region, the additional term $V_{\mathrm{B}}=V_{0}\hat{\tau}_{z}$
opens an insulating gap of size $2V_{0}$, which accounts for contact deficiencies between
the pump and electrode.

The Hamiltonian in the pump is Dirac-like, while in the metal electrode, the Hamiltonian
is parabolic. It is well-known that the wavefunction of a parabolic Hamiltonian can not
be connected directly to that of a Dirac-like Hamiltonian. To overcome this problem,
following Chen $et$ $al.$\cite{chen}, we linearize the Hamiltonian Eq.\ (\ref{H_e}) around
the Fermi energy before proceeding. When $E_{0}$ is sufficiently large, for a given $p_{y}$,
we can linearize the effective 1D Hamiltonian $H_{\mathrm{E}}$ at the right and left Fermi
points $p_{x}=\pm mv^{\prime }_{\mbox{\tiny F}}(k_y)$ with
$v^{\prime }_{\mbox{\tiny F}}(k_y)=\sqrt{2m(E_{\mbox{\tiny F}}+E_{0})-k_{y}^2}/m$. A Pauli
matrix $\hat{\tau}_x$ is introduced to describe the right and left-moving branches. To be
consistent with the form of the Hamiltonian of the pump and also preserve the time-reversal
symmetry, we use $\tau_{x}=1$ and $-1$, respectively, to represent the right-moving and
left-moving branches for $\eta=1$, and oppositely for $\eta=-1$. As a result, the
Hamiltonian of the electrode becomes
\begin{equation}
H_{\mathrm{E}}=v^{\prime }_{\mbox{\tiny F}}k_{x}\eta\hat{\tau}_{x}\ ,
\label{H_e1}
\end{equation}
where $k_y=p_{y}$ and $k_x=p_{x}\mp mv^{\prime }_{\mbox{\tiny F}}(k_y)$ for the right and
left-moving branches. Strictly speaking, the operator $\hat{\tau}_x$ and the momentum
$\mathbf{k}$ in the electrode have different physical meaning from those in the pump. We
will omit such a difference, assuming that this mismatch can be effectively accounted by
the potential barrier. The pumping effect is usually dominated by small $k_{y}$, so that
we can further approximate $v^{\prime }_{\mbox {\tiny F}}(k_{y})\simeq
v^{\prime }_{\mbox {\tiny F}}(k_{y}=0) \equiv v^{\prime }_{\mbox {\tiny F}}$, with purpose
to minimize the number of adjustable parameters in the model.

Calculation of the number of electrons pumped per cycle amounts to solving the scattering
problem for an electron at the Fermi energy incident from the electrode. The Fermi energy
will be set to be $E_{\mbox{\tiny F}}=0$, which is in the band gap of the pump. In this
case, the incident electron will be fully reflected back into the electrode. In order to
obtain the scattering amplitudes, we need to solve the wavefunctions in the three regions.
For a spin $\sigma$ electron incident from $\eta$ valley, the wavefunction in the electrode
is given by
\begin{equation}
\Psi^{\eta\sigma}_{\mathrm{E}}(x)=\frac{1}{\sqrt{2}}\left(
\begin{array}{c}
1 \\
-\eta \\
\end{array}
\right)+ \frac{r_{\eta}^{\sigma}(k_{y})}{\sqrt{2}}\left(
\begin{array}{c}
1 \\
\eta \\
\end{array}
\right)\ .  \label{Wave_E}
\end{equation}
The wavefunctions in the potential barrier and in the pump can be written as
\begin{equation}
\Psi^{\eta\sigma}_{\mathrm{B}}(x)=C_{\eta1}\left(
\begin{array}{c}
1 \\
\eta i \\
\end{array}
\right)e^{-\gamma_{0}x}+ C_{\eta2}\left(
\begin{array}{c}
1 \\
-\eta i \\
\end{array}
\right)e^{\gamma_{0}x}\ ,  \label{Wave_Barrier}
\end{equation}
\begin{equation}
\Psi^{\eta\sigma}_{\mathrm{P}}(x)=t_{\eta}^{\sigma}(k_{y})\left(
\begin{array}{c}
i\sin{\frac{\varphi_{\eta}^{\sigma}}{2}} \\
\cos{\frac{\varphi_{\eta}^{\sigma}}{2}} \\
\end{array}
\right)e^{k_{x}x},  \label{Wave_Pump}
\end{equation}
where $\gamma_{0}=V_{0}/\hbar v^{\prime }_{\mbox{\tiny F}}$,
$k_{x}=\sqrt{{x_{\eta}^{\sigma}}^{2}+{y_{\eta}^{\sigma}}^{2}}/v_{\mbox{\tiny F}}$,
$\varphi_{\eta}^{\sigma}=\mathrm{arg}[x_{\eta}^{\sigma} +iy_{\eta}^{\sigma}]$
with $x_{\eta}^{\sigma}=-\eta v_{\mbox{\tiny F}}[k_{y}-eA_{y}\sin{(\omega t})]$ and $y_{\eta}^{\sigma}=\eta[\eta\sigma\lambda_{\mathrm{SO}}-lE_{z}^{0}-lE_{z}^{1}\cos{(\omega t)}]$.
Matching the wavefunctions given in Eqs.\ (\ref{Wave_E}), (\ref{Wave_Barrier}) and
(\ref{Wave_Pump}) at $x=0$ and $x=d$ by using the boundary conditions
$\Psi^{\eta\sigma}_{\mathrm{E}}(d+0^{+})=\Psi^{\eta\sigma}_{\mathrm{B}}(d-0^{+})$ and
$\Psi^{\eta\sigma}_{\mathrm{B}}(0^{+})=\Psi^{\eta\sigma}_{\mathrm{P}}(0^{-})$, we can
obtain for the reflection amplitudes
\begin{equation}
r_{\eta}^{\sigma}(k_{y})=-\frac{\cos{\varphi_{\eta}^{\sigma}} +i[\mbox{sh}(2\gamma_{0}d)+\eta\sin{\varphi_{\eta}^{\sigma}}\mbox{ch}(2\gamma_{0}d)]}
{\mbox{ch}(2\gamma_{0}d) +\eta\sin{\varphi_{\eta}^{\sigma}}\mbox{sh}
(2\gamma_{0}d)}\ .  \label{reflection amplitude}
\end{equation}

The number of electrons of valley $\eta$ and spin $\sigma$ pumped per cycle
at momentum $k_{y}$ is given by~\cite{butti,brou}
\begin{equation}
\Delta n_{\eta}^{\sigma}(k_{y})=\frac{1}{2\pi i}\oint_{T}{r_{\eta}^{\sigma}}%
^{*}(k_{y}) dr_{\eta}^{\sigma}(k_{y})\ .  \label{Electrons_Pumped}
\end{equation}
Noting that $\vert r_{\eta}^{\sigma}(k_{y})\vert \equiv 1$ due to full reflection, one
can easily identify $\Delta n_{\eta}^{\sigma}(k_{y})$ with the winding number of
$r_{\eta}^{\sigma}(k_{y})$ around the origin on the complex plane in a cycle.
$\vert r_{\eta}^{\sigma}(k_{y})\vert \equiv 1$ also indicates that with changing the
barrier strength $\gamma_{0}d$, the trajectory of $r_{\eta}^{\sigma}(k_{y})$ will never
sweep through the origin, and so the winding number is invariable. As a result,
$\Delta n_{\eta}^{\sigma}(k_{y})$ is independent of the barrier strength $\gamma_{0}d$.
Thus, we can calculate $\Delta n_{\eta}^{\sigma}(k_{y})$ simply by setting $\gamma_{0}d=0$,
and the result is valid for any barrier strength. For $\gamma_{0}d=0$,
$r_{\eta}^{\sigma}(k_{y})=-(\cos{\varphi_{\eta}^{\sigma}} +i\eta\sin{\varphi_{\eta}^{\sigma}})=e^{i(\pi+\eta\varphi_{\eta}^{\sigma})}$,
and we can derive Eq.\ (\ref{Electrons_Pumped}) to be
$\Delta n_{\eta}^{\sigma}(k_{y})=\eta[\varphi_{\eta}^{\sigma}(T)-\varphi_{\eta}^{\sigma}(0)]/2\pi$.

Because of the periodicity, the increment of $\varphi_{\eta}^{\sigma}(t)$
 in a period, namely, $\varphi_{\eta}^{\sigma}(T)-\varphi_{\eta}^{\sigma}(0)$,
 must be integer multiples of $2\pi$. From the
expression for $\varphi_{\eta}^{\sigma}(t)$ given below Eq.\ (\ref{Wave_Pump}), we know that
$\varphi_{\eta}^{\sigma}(t)$ is the argument of $x_{\eta}^{\sigma}+iy_{\eta}^{\sigma}$. It is
clear that the trajectory of $x_{\eta}^{\sigma}+iy_{\eta}^{\sigma}$ is an ellipse on the
complex plane centered at
$[-\eta v_{\mbox{\tiny F}}k_{y}, -\eta(lE_{z}^{0}-\eta\xi_{\sigma}\lambda_{\mathrm{SO}})]$,
with $|ev_{\mbox{\tiny F}} A_{y}|$ and $|lE_{z}^{1}|$ as the semi-major and semi-minor axes
oriented along the real and imaginary axes. If the ellipse encircles the origin $(0,0)$, the
increment of $\varphi_{\eta}^{\sigma}(t)$ takes value $2\pi$ or $-2\pi$, depending on the
direction of the trajectory. Otherwise, the increment is $0$. The direction of the trajectory
is determined by the sign of $E_{y}E_{z}^{1}$. For the ellipse of
$x_{\eta}^{\sigma}+iy_{\eta}^{\sigma}$ to surround the origin, two sufficient and necessary
conditions must be satisfied. First, the ellipse needs to intersect the real axis. This
requires that the maximum and minimum of $y_{\eta}^{\sigma}$ have opposite signs, and so
$|lE_{z}^{1}|>|lE_{z}^{0}-\lambda_{\mathrm{SO}}|$ for $\eta\xi_{\sigma}=1$,
and $|lE_{z}^{1}|>|lE_{z}^{0}+\lambda_{\mathrm{SO}}|$ for $\eta\xi_{\sigma}=-1$. Second, the
two intersecting points need to be located at opposite sides of the origin. This results in
the condition $|k_{y}|<k_{y}^{c+}$ for $\eta\xi_{\sigma}=1$, and $|k_{y}|<k_{y}^{c-}$ for
$\eta\xi_{\sigma}=-1$. Now it is easy to see that the number of electrons
for given valley $\eta$ and spin $\sigma$ pumped per cycle at momentum $k_{y}$
equals to the spin-valley Chern number $\Delta n_{\eta}^{\sigma}(k_{y})=C_{\eta}^{\sigma}(k_{y})$.

\begin{figure}[tpb]
\begin{center}
\includegraphics[width=5.0in,keepaspectratio]{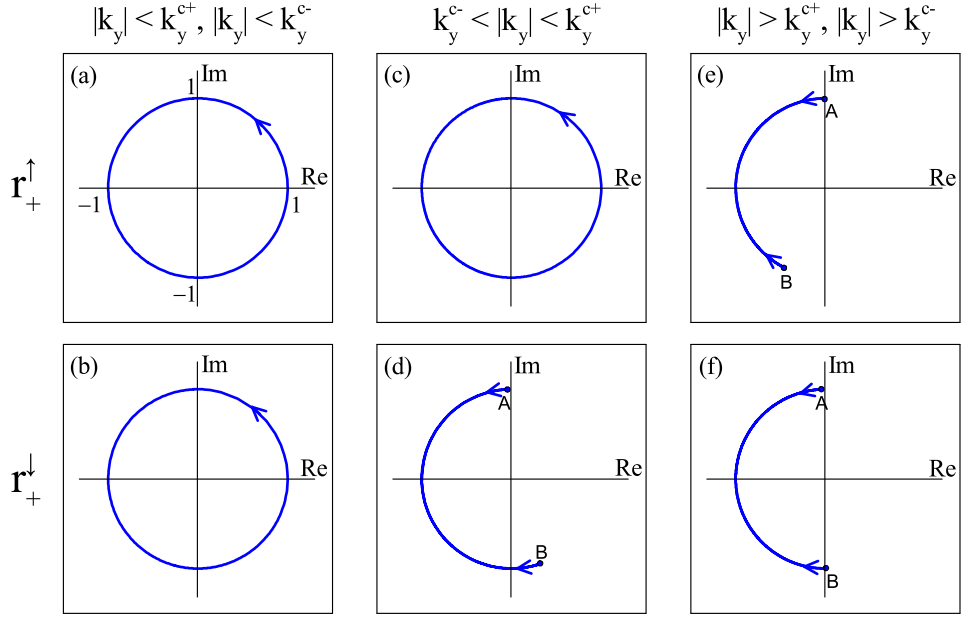}
\caption{Trajectories of the reflection amplitudes in a cycle on the complex plane,
in which only the $\eta=+$ valley is considered.
The upper and lower rows correspond to spin $\sigma=\uparrow$ and $\downarrow$,
respectively.
 In (a) and (b), $k_{y}=0.5eA_{y}$,
in (c) and (d), $k_{y}=0.8eA_{y}$,
and in (e) and (f), $k_{y}=eA_{y}$.
The other parameters are set to be $\gamma_{0}d=1$,
$lE_{z}^{0}=0.5\lambda_{\mathrm{SO}}$
and $lE_{z}^{1}=2\lambda_{\mathrm{SO}}$,
for which the corresponding $k_{y}^{c+}$ and $k_{y}^{c-}$ equal to
$\frac{\sqrt{15}}{4}eA_{y}$ and $\frac{\sqrt{7}}{4}eA_{y}$, respectively.
In (d)(e)(f), the trajectories start from point $A$, travel to $B$,
and then return.
}\label{Fig.4}
\end{center}
\end{figure}

To further confirm the above general discussion, in Figs.\ 3(a)-3(f), we plot the
trajectories of $r_{\eta}^{\sigma}(k_{y})$ for the $\eta=+$ valley for momentum $k_y$
in different regions. In (a) and (b), the conditions $\vert k_{y}\vert<k_{y}^{c+}$
and $\vert k_{y}\vert<k_{y}^{c-}$ are satisfied, corresponding to the yellow region
in Fig.\ 2(b), and both $r_{+}^{\uparrow}(k_{y})$ and $r_{+}^{\downarrow}(k_{y})$
go around the origin counterclockwise once in a cycle. As a result,
$\Delta n_{+}^{\uparrow}(k_{y})=\Delta n_{+}^{\downarrow}(k_{y}) =1$, in agreement
with the spin-valley Chern numbers $C_{+}^{\uparrow}(k_{y})=C_{+}^{\downarrow}(k_{y})=1$.
In (c) and (d), we have $k_{y}^{c-}<\vert k_{y}\vert<k_{y}^{c+}$, corresponding to
the right blue region in Fig.\ 2(b), and $r_{+}^{\uparrow}(k_{y})$ goes around the
origin once, but $r_{+}^{\downarrow}(k_{y})$ does not. Therefore,
$\Delta n_{+}^{\uparrow}(k_{y})= 1$ and $\Delta n_{+}^{\downarrow}(k_{y})= 0$, in
agreement with $C_{+}^{\uparrow}(k_{y})=1$ and $C_{+}^{\downarrow}(k_{y})=0$. In (e)
and (f), we have $\vert k_{y}\vert>k_{y}^{c+}$ and $\vert k_{y}\vert>k_{y}^{c-}$,
corresponding to the white region in Fig.\ 2(b), and the winding numbers of
$r_{+}^{\uparrow}(k_{y})$ and $r_{+}^{\downarrow}(k_{y})$ around the origin are zero.
Therefore, $\Delta n_{+}^{\uparrow}(k_{y})=\Delta n_{+}^{\downarrow}(k_{y})= 0$, in
agreement with $C_{+}^{\uparrow}(k_{y})=C_{+}^{\downarrow}(k_{y})=0$. The calculated
trajectories of $r_{+}^{\uparrow}(k_{y})$ and $r_{+}^{\downarrow}(k_{y})$ are fully
consistent with the spin-valley Chern number description.

Based upon the above discussions or directly from the expressions for the spin-valley
Chern numbers, Eqs.\ (\ref{spin-valley1}) and (\ref{spin-valley2}), one can obtain for
the total spin pumped per cycle by electron states of all momenta
\begin{equation}
\Delta S=\hbar\frac{L_{y}}{\pi}\mathrm{sgn}(E_{y}E_{z}^{1})\Bigl
[\theta(|lE_{z}^{1}|- |lE_{z}^{0}-\lambda_{\mathrm{SO}}|)k_{y}^{c+}
-\theta(|lE_{z}^{1}|-|lE_{z}^{0}+\lambda_{\mathrm{SO}}|)k_{y}^{c-}\Bigr]\ .
\label{total spin Pumped}
\end{equation}
$\Delta S$ is in scale with the width $L_{y}$ of the pump, a clear indication that the
spin pumping is a bulk effect. If $E_{z}^{0}=0$, where $k_{y}^{c+}=k_{y}^{c-}$, we have
$\Delta S=0$. As discussed earlier, the system is in the pure valley pumping regime,
without pumping spin. If $|lE_{z}^{1}|< |lE_{z}^{0}-\lambda_{\mathrm{SO}}|$ and
$|lE_{z}^{1}|<|lE_{z}^{0}+\lambda_{\mathrm{SO}}|$, which corresponds to the
trivial insulator phase in the white region of Fig.\ 1(a), we also have $\Delta S=0$. In
all other cases, $\Delta S\neq 0$, and the system serves as a topological spin Chern pump.

\section*{Conclusion}

We have investigated the topological pumping effect in silicene, modulated
by an in-plane and a vertical time-dependent electric field. Using
spin-valley Chern numbers to characterize the topological pumping, we find
that there exist three quantum pumping regimes in the system, a pure valley
pumping regime, a spin-valley pumping regime, and a trivial insulator
regime, depending on the strengths of the electrostatic and $ac$ components
of the perpendicular electric field. The amount of spin pumped per cycle
calculated from the scattering matrix formula is fully consistent with the
topological description based upon the spin-valley Chern numbers. This work
proposed a relatively easy scheme to achieve topological spin or valley
Chern pumping. It also demonstrates the fact that bulk topological spin or
valley Chern pumping is a characteristic observable effect of various QSH
systems, if the material parameters of the QSH systems can be suitably
modified with time.

\section*{Acknowledgments}
This work was supported by the State Key Program for Basic Researches of
China under grants numbers 2015CB921202, 2014CB921103 (LS), the National
Natural Science Foundation of China under grant numbers 11225420 (LS),
11174125, 91021003 (DYX) and a project funded by the PAPD of Jiangsu Higher
Education Institutions.

\section*{Author contributions}
W.L. carried out the numerical calculations. W.L. and L.S. analyzed the results.
L.S., B.G.W. and D.Y.X. guided the overall project. All authors reviewed the
manuscript. All authors participated in discussions and approved the submitted
manuscript.

\section*{Additional information}
\textbf{Competing financial interests:} The authors declare no competing financial interests.

\end{document}